# Magnon-mediated perpendicular magnetization switching by topological crystalline insulator SnTe with high spin Hall conductivity

Pengnan Zhao[1,#], Guoyi Shi[2,#], Wentian Lu[1,#], Lihuan Yang[1], Hui Ru Tan[3], Kaiwei Guo[1], Jia-Min Lai[1], Zhonghai Yu[1], Anjan Soumyanarayanan[3,4], Zhe Yuan[5,6], Fei Wang[1,*], Xiaohong Xu[1,†], and Hyunsoo Yang[2,‡]

[1]Key Laboratory of Magnetic Molecules and Magnetic Information Materials of Ministry of Education & School of Chemistry and Materials Science of Shanxi Normal University, Taiyuan 030006, China

[2]Department of Electrical and Computer Engineering, National University of Singapore, Singapore, 117576, Singapore

[3]Institute of Materials Research & Engineering, Agency for Science, Technology & Research (A*STAR), Singapore 138634, Singapore

[4]Department of Physics, National University of Singapore, Singapore 117551, Singapore

[5]Institute for Nanoelectronic Devices and Quantum Computing, Fudan University, Shanghai 200433, China

[6]Interdisciplinary Center for Theoretical Physics and Information Sciences, Fudan University, Shanghai 200433, China

*Contact author: feiwang.imr@gmail.com
†Contact author: xuxh@sxnu.edu.cn
‡Contact author: eleyang@nus.edu.sg

**ABSTRACT**: Magnons possess the ability to transport spin angular momentum in insulating magnetic materials, a characteristic that sets them apart from traditional electronics where power consumption arises from the movement of electrons. However, the practical application of magnon devices demands room temperature operation and low switching power of perpendicular magnetization. Here we demonstrate the low-power manipulation of perpendicular magnetization via magnon torques in SnTe/NiO/CoFeB devices at room temperature. Topological crystalline insulator SnTe exhibits a high spin Hall conductivity of $\sigma_s \approx 6.1 \times 10^4$ ($\hbar/2e$) $(\Omega \cdot m)^{-1}$, which facilitates the generation of magnon currents in an antiferromagnetic insulator NiO. The magnon currents traverse the 20-nm-thick NiO layer and subsequently exert magnon torques on the adjacent ferromagnetic layer, leading to magnetization switching. Notably, we achieve a 22-fold reduction in power consumption in SnTe/NiO/CoFeB heterostructures compared to $Bi_2Te_3$/NiO/CoFeB control samples. Our findings establish the low-power perpendicular magnetization manipulation through magnon torques, significantly expanding the range of topological materials with practical applications.

## I. INTRODUCTION

A magnon is a quasiparticle that embodies the quantized excitations of spin waves in a magnetic material[1-5]. When magnons collectively migrate from one region to another, their coordinated movement gives rise to a magnon current. Unlike an electric current, which entails the flow of charge carriers within conductive materials, magnon currents can transport spin angular momentum through magnetic insulators[1,5]. This characteristic of magnon currents avoids the Joule heating associated with electron movement in electronic systems, thereby offering a low-power method for the transfer of spin and energy in spintronic applications. Additionally, magnon transmission in insulating materials can span micrometer-scale distances[6-10] at ultrafast velocities[11]. These distinctive features of magnons lay the groundwork for the development of high-speed and energy-efficient spintronic devices.

Magnons can exert spin-transfer torques on local magnetic moments[12], thereby inducing the motion of domain walls[13-16] or magnetization reversal[13-16]. Magnon-mediated switching of ferromagnetic magnetization has been demonstrated in the $Bi_2Se_3$/NiO/Py[17], Pt/NiO/$Y_3Fe_5O_{12}$[18], $SrIrO_3$/NiO/$SrRuO_3$[19,20], $Bi_2Te_3$/NiO/CoFeB[21], and $SrIrO_3$/$BiFeO_3$/$SrRuO_3$[22] systems. It is important to note that the magnon torque efficiency is highly sensitive to temperature[20], and the magnetization switching generally vanishes at room temperature, with a few exceptions reported for the $Bi_2Se_3$/NiO/Py[17] and $Bi_2Te_3$/NiO/CoFeB[21] systems. However, topological insulators ($Bi_2Se_3$ and $Bi_2Te_3$) suffer from a high resistivity [21], leading to significant power consumption issues. Therefore, the primary challenge for magnon device applications is to achieve both room temperature operation and low-power switching. In this work, we utilize a topological crystalline insulator SnTe/NiO/ferromagnet sandwich heterostructure. Compared to the $Bi_2Te_3$-based system, SnTe achieves a 22-fold reduction in power consumption for magnon-mediated switching at room temperature. This significant reduction in power consumption is attributed to the high spin Hall conductivity of SnTe.

## II. EXPERIMENTAL SECTION

### A. Sample Preparation.

SnTe films are grown in a commercial molecular beam epitaxy (MBE) chamber (Prevac systems) with a base pressure of $\sim 1\times 10^{-10}$ mbar. A 1 nm insulating $Bi_2Te_3$ layer is grown on the heat-treated $Al_2O_3$ (001) substrates at 200 °C as a buffer layer. High-purity Sn (99.999%) and Te (99.9999%) are evaporated from Knudsen effusion cells. During the SnTe film growth, the substrate is maintained at ~ 220 °C. The flux ratio of Te/Sn is set to be ~ 10 and the SnTe growth rate is ~ 0.2 nm/min. Following the growth, the SnTe films are annealed at ~250 °C for 30 minutes to improve the crystal quality before being cooled down to room temperature. The SnTe films are immediately transferred to a magnetron sputtering chamber. An antiferromagnetic NiO layer with different thicknesses is sputtered onto the SnTe films at room temperature. NiO exhibits a polycrystalline morphological texture. Next, Ti (2 nm)/$Co_{0.2}Fe_{0.6}B_{0.2}$ (0.9 nm)/MgO (2 nm)/Ta (1.5 nm) for perpendicular magnetization switching measurements and $Ni_{81}Fe_{19}$ (6 nm)/$SiO_2$ (2 nm)/Ta (1.5 nm) for spin-torque ferromagnetic resonance (ST-FMR) measurements are deposited onto the NiO layer at room temperature. The Ta layer is oxidized upon venting the growth chamber.

### B. ST-FMR Devices Fabrications and Measurements.

The films are patterned into rectangular microstrips with a width of 30 μm and length of 10 μm using photolithography and ion milling techniques. Subsequently, Ta/Pt electrodes are deposited onto the devices using the magnetron sputtering. We employ the ST-FMR technique to quantify the spin-orbit torques generated in the SnTe/NiO/Py systems. An in-plane radio frequency current ($I_{RF}$) with frequencies ranging from 7 to 12 GHz and a power of 15 dBm is applied with an in-plane external magnetic field. This $I_{RF}$ induces oscillating spin currents in the SnTe layer, which are subsequently converted into magnon currents through the interfacial exchange interaction between SnTe and NiO. These magnon currents propagate through the NiO layer, generating oscillating magnon torques on the adjacent Py layer. These torques include both the damping-like torque ($\bm{m} \times \bm{\sigma} \times \bm{m}$) and the field-like torque ($\bm{m} \times \bm{\sigma}$), where $\bm{\sigma}$ presents the induced magnon spin polarization in the NiO layer. Moreover, the $I_{RF}$ induces the Oersted field torque $-\gamma\,(\bm{m} \times \bm{H_{Oe}})$ on the Py layer. The combined torques, which can be decomposed into out-of-plane and in-plane components, drive the magnetization of the Py layer away from its equilibrium and into precession, causing a change in the anisotropic magnetoresistance. Consequently, the interaction between the varying device resistance and $I_{RF}$ produces a direct current (d.c.) voltage, measured as the ST-FMR signal $V_{mix}$ by a lock-in amplifier.

### C. Hall Bar Device Fabrication and Switching Measurements.

The films are patterned into 40 × 20 μm Hall bars for switching measurements using a combination of optical lithography and ion milling techniques. Subsequently, Ta/Pt electrodes are deposited onto the devices using magnetron sputtering. The electrical transport measurements are conducted using a Quantum Design Physical Property Measurement System (PPMS). The Keithley 6221 serves as the current source for both d.c. and pulse measurements. During the pulse measurements, a 100 μs pulse is applied, followed by the application of a small d.c. current of 50 μA to detect the Hall resistance. The Hall voltage is then determined using a Keithley 2182A nanovoltmeter.

### III. RESULTS AND DISCUSSION

SnTe is a topological crystalline insulator characterized by its topological properties that are protected by crystal symmetry[23,24]. Unlike topological insulators such as $Bi_2Te_3$, which possess a single Dirac cone, SnTe possesses four Dirac surface states and is therefore expected to have a higher spin Hall conductivity[25,26]. We grow SnTe films with different thicknesses ($d$ = 3, 4, 5, 6, 7, 8, 9, 10, 15 and 20 nm) on sapphire (001) substrates using molecular beam epitaxy. The high crystalline quality of the SnTe films is confirmed through reflection high-energy electron diffraction, Raman spectroscopy and X-ray diffraction (Section S1, see Supplemental Material [27]). The subsequent growth is carried out in a magnetron sputtering chamber for the fabrication of SnTe ($d$)/NiO ($t$)/$Ni_{81}Fe_{19}$ (6)/$SiO_2$ (2)/Ta (1.5) (henceforth referred to as SnTe ($d$)/NiO ($t$)/Py, with the thicknesses in nanometers indicated in parentheses), where the Ta layer forms insulating $TaO_x$ upon air exposure. Subsequently, SnTe ($d$)/NiO ($t$)/Py heterostructures are fabricated into ST-FMR devices using a combination of optical lithography and ion milling techniques[28]. The Py layer in all devices exhibits in-plane magnetic anisotropy.

Figure 1a depicts representative ST-FMR signals $V_{mix}$, for the SnTe (8 nm)/Py sample. The $V_{mix}$ signal can be decomposed into $V_{mix} = V_S F_S + V_A F_A$, where the symmetric component, $V_S F_S$, is attributed to the in-plane anti-damping torque, and the antisymmetric component, $V_A F_A$, arises from a combination of the field-like torque and the Oersted field torque. By employing a well-established analysis procedure[28-30], the effective spin-orbit torque efficiency ($\theta_y$) is extracted to be = 0.165. In addition, we perform ST-FMR measurements across SnTe ($d$)/Py samples with varying $d$, and the results are presented in Figure 1b. The data exhibit a monotonic increase in $\theta_y$ with increasing $d$, reaching saturation at $d$ = 8 nm.

The spin diffusion length is fitted to be 4.1 nm, comparable to that of WTe$_2$ (~ 4.3 nm) and PtTe$_2$ (~ 5.1 nm)[28] (Section S2, see Supplemental Material [27]).

The room temperature resistivity ($\rho_{xx}$) of the 8 nm SnTe film is 270 μΩcm and thus the spin Hall conductivity (SHC, $\sigma_s$) is determined to be 6.1 × 10$^4$ (ℏ/2e) (Ωm)$^{-1}$, which is 3 times higher than that of Bi$_2$Te$_3$[21]. The SHC observed in SnTe can be understood through *ab initio* calculations based on density functional theory. Figure 1c shows the calculated electronic structure of prototypical SnTe, confirming its character of a topological crystalline insulator with a band gap of ~ 0.1 eV. Utilizing the maximally localized Wannier functions (MLWFs) method under a frozen disentanglement window, we are able to accurately reproduce the *ab initio* band structure, which validates the use of Wannier interpolation for precise SHC calculations[31]. We then calculate the intrinsic SHC of SnTe as a function of energy using *ab initio* methods, where the applied electrical field is along the (111) crystallographic direction. Due to the presence of a small band gap, we set the SHC at the Fermi level to zero to maintain consistency with the electronic band results. Below the Fermi level ($E − E_F < 0$), the SHC exhibits very weak energy dependence, as depicted in Figure 1d. At $E − E_F > 0$, the SHC increases with shifting the chemical potential into conduction bands with a peak of 1.4 × 10$^5$ (ℏ/2e) (Ωm)$^{-1}$ at $E_F$ + 0.7 eV, indicating a significant enhancement in SHC for electron-doped carriers. Because our SnTe samples exhibit heavy hole doping (Section S1, see Supplemental Material [27]), the SHC for SnTe is estimated to be 3 × 10$^4$ (ℏ/2e) (Ωm)$^{-1}$, which is about half of the experimentally observed value of $\sigma_s$ = 6.1 × 10$^4$ (ℏ/2e) (Ωm)$^{-1}$. We can attribute this discrepancy to extrinsic contributions, such as side-jump and skew-scattering effects[32], which is not accounted for in our calculations. To elucidate the underlying mechanism of SHC variation, we plot *k*-resolved spin Berry curvature term in a two-dimensional Brillouin zone. As shown in Figure 1e, positive contributions dominate in most regions of the Brillouin zone, with the Γ point playing a significant role. Moreover, we depict the spin Berry curvature term at $E = E_F + 0.7$ eV (Section S3, see Supplemental Material [27]). At this energy, the SHC is enhanced by at least 4 times compared to the value when the chemical potential is located in the valence bands, suggesting that SHC can be significantly improved by tuning the chemical potential to conduction bands.

Figure 2a depicts the representative $V_{mix}$ signals for the SnTe (8 nm)/NiO (20 nm)/Py sample, from which $\theta_y$ is calculated to be 0.083. Additionally, we perform ST-FMR measurements with varying thicknesses (*t*) of NiO in SnTe (8 nm)/NiO (*t*)/Py samples. The summarized results are presented in Figure 2b. The data demonstrate a noticeable decline in $\theta_y$ as the NiO thickness increases from $t = 0$ to 3 nm. This decrease is attributed to the suppression of electron-mediated spin currents by the non-magnetic NiO layer[17]. Beyond $t = 3$ nm, the amplitude of $\theta_y$ progressively increases, with a peak at $t = 20$ nm, indicating the establishment of antiferromagnetic ordering within the NiO layer (Section S4, see Supplemental Material [27]). However, for $t > 20$ nm, the efficiency of the magnon torque diminishes, primarily due to increased spin angular momentum dissipation.

We investigate whether the observed magnon torques can switch the perpendicular magnetization at room temperature. A film stack of Ti (2 nm)/$Co_{20}Fe_{60}B_{20}$ (0.9 nm)/MgO (2 nm)/Ta (1.5 nm) (abbreviated CoFeB) is deposited on top of SnTe (8 nm)/NiO (20 nm) using magnetron sputtering. The choice of a 20 nm NiO is based on the thickness at which the magnon torque efficiency reaches its maximum in Figure 2b. The SnTe/NiO/CoFeB films are patterned into Hall bar devices with dimensions of 40 μm in the length and 20 μm in the width. The anomalous Hall measurement displays square hysteresis loops, indicating long-range ferromagnetic order with perpendicular anisotropy of CoFeB in Figure 3a. Pulsed direct currents (d.c.) are applied to the Hall bar devices, and the corresponding Hall resistance is measured. A clear switching window is observed when an external magnetic field ($\mu_0 H_x$) of 10 mT is applied along the current direction (Figure 3b). The critical switching current density ($J_C$) is determined to be $5.46 \times 10^6$ A/cm$^2$, which is lower than the values typically observed in conventional electron-mediated systems ($J_C \sim 10^7$ A/cm$^2$)[33-35]. The ratio of the change in the switching resistance to the anomalous Hall resistance ($R_S/R_H$) is ~ 0.72 (Figure 3c). The switching of the CoFeB layer suggests the existence of in-plane anti-damping magnon torques. We calculate the power consumption of magnon devices based on a two current model[36]. The power consumption of the SnTe (8 nm)/NiO (20 nm)/CoFeB sample is 22 times smaller than the control sample of Bi$_2$Te$_3$ (8 nm)/NiO (25 nm)/CoFeB (Section S5 and S6, see Supplemental Material [27]), indicating a low-power manipulation of SnTe-based magnon devices at room temperature. The temperature rise during magnetization switching measurements in the SnTe (8 nm)/NiO (20 nm)/CoFeB sample is evaluated to be 11.3 K (Section S7, see Supplemental Material [27]).

With decreasing $\mu_0 H_x$, $R_S/R_H$ decreases, and $J_C$ increases. At $\mu_0 H_x = 0$ mT, the switching window vanishes, indicating that the magnon-mediated switching of perpendicular magnetization necessitates the assistance of an external magnetic field. As $\mu_0 H_x$ is further decreased to a negative value, the switching loop gradually recovers, and the switching polarity changes from clockwise to anticlockwise. Figure 3d presents the switching phase diagram, revealing that $J_C$ slightly decreases with increasing $\mu_0 H_x$. This behavior is consistent with that observed in other systems[21].

Figure 4 provides a summary of the $t$ dependence of $R_S/R_H$ (Figure 4a) and $J_C$ (Figure 4b) in SnTe (8 nm)/NiO ($t$)/CoFeB heterostructures under an external magnetic field of 10 mT. At $t = 0$, a $R_S/R_H$ ratio reaches its maximum, and the $J_C$ value is minimized, in which the torques are primarily driven by itinerant electron spins (Section S8, see Supplemental Material [27]). Within the $t$ range from 3 to 10 nm, the switching is suppressed, attributed to the attenuation of spin currents and their negligible spin-orbit torques (Section S9, see Supplemental Material [27]). When $t$ exceeds 10 nm, the switching window re-emerges, with a peak of $R_S/R_H$ and a local minimum in $J_C$ observed at $t = 20$ nm, correlating with the highest magnon torque efficiency. Beyond $t = 20$ nm, the switching efficiency declines, consistent with the results of the ST-FMR measurements in Figure 2b. Moreover, we note that the maximum magnon efficiency in topological material-based magnon devices is achieved when the NiO layer has a thickness of 20 to 25 nm[11,17,21], whereas for Pt-based magnon devices, the peak efficiency occurs with an a NiO thickness of 0.5 to 2 nm[11,18,37,38]. This discrepancy can be attributed to the impact of the underlying spin source layer materials on the anisotropy and exchange interaction in NiO, leading to modifications in spin transport properties (Section S10, see Supplemental Material [27]).

We now explain the magnon torque-driven switching in the SnTe/NiO/CoFeB heterostructures. When a charge current flows through the SnTe layer, it induces spin accumulation with the polarization **σ** at the SnTe/NiO interface. The spin accumulation can excite antiferromagnetic magnon modes depending on the antiferromagnetic Néel vector **L** of NiO[21,39,40], as the generation and transport of magnons are proportional to $\boldsymbol{\sigma}\cdot\boldsymbol{L}$.[8] In the case of a polycrystalline antiferromagnet NiO (Section S1, see Supplemental Material [27]), $\boldsymbol{L}$ is approximately uniformly distributed, allowing magnon currents in any polarization direction to transmit the polycrystalline NiO layer. Upon reaching the CoFeB layer, these magnons are absorbed, exerting magnon torques on the local magnetic moments. The in-plane anti-damping magnon torques facilitate deterministic switching of the CoFeB layer with the assistance of an external magnetic field.

In the SnTe/NiO/CoFeB heterostructures, magnon currents are triggered by a charge current flowing through the spin source layer, SnTe. The conversion and transmission of magnon currents inherently lead to the loss of spin angular momentum, which consequently causes the power consumption of magnon torques in SnTe/NiO/CoFeB systems to be higher than that of electron-mediated torques in SnTe/CoFeB systems (Section S11, see Supplemental Material [27]). However, it is important to note that magnon-based devices are still in their infancy. The information writing in magnon devices can be electrically isolated[17,21], which addresses technical hurdles related to vias and interconnects. In spin source layer/antiferromagnetic insulator/ferromagnet trilayers, the magnon current is excited through spin accumulation at the spin source layer/antiferromagnet interface. Effective magnon excitation requires the spin source material to exhibit a sufficiently large spin torque efficiency to generate adequate spin accumulation. This requirement limits the number of systems capable of demonstrating magnon torque-driven magnetization switching at room temperature (Section S11, see Supplemental Material [27]). Our work is significant in this context, as it not only introduces a new addition to the limited group of room-temperature magnon devices, but more importantly, it demonstrates the low-power switching of perpendicular magnetization through magnon torques. Further reduction of the power consumption can be achieved by optimizing the Néel vector orientation of the utilized antiferromagnet along the σ direction. Moreover, we anticipate the creation of devices that operate exclusively on magnons, eliminating the need for charge currents. In this sense, magnetoelectric multiferroic materials may be utilized by applying pulsed voltages, thereby generating magnons with significantly lower power consumption[41].

## IV. CONCLUSIONS

We have demonstrated magnon-mediated switching of perpendicular magnetization in SnTe/NiO/CoFeB heterostructures. Previously, magnon torque-based magnetization switching at room temperature was demonstrated using topological insulators, leading to significant power consumption due to a high resistivity of topological insulators. In this work, we utilize a topological crystalline insulator SnTe, and the power consumption of our magnon devices is 22 times lower compared to those utilizing the topological insulator $Bi_2Te_3$ as the magnon current generator. The substantial reduction in power consumption is primarily attributed to the high spin Hall conductivity of SnTe, as confirmed by both ST-FMR measurements and theoretical calculations. Our research establishes a robust foundation for the development of high-density and low-power-consumption magnon devices.

## ACKNOWLEDGMENTS

H.Y. is supported by National Research Foundation (NRF) Singapore Investigatorship (NRFI06-2020-0015). F. W. and X. X. acknowledge the support from National Key Research and Development Program of China (2022YFB3505301), the National Natural Science Foundation of China (12174237, 52471253), Fund Program for the Scientific Activities of Selected Returned Overseas Professionals in Shanxi Province (20240019), and Central Government's Special Fund for Local Science and Technology Development (YDZJSX2024D058). W. L. acknowledges the financial support by the Basic Research Plan of Shanxi Province (202203021212393).

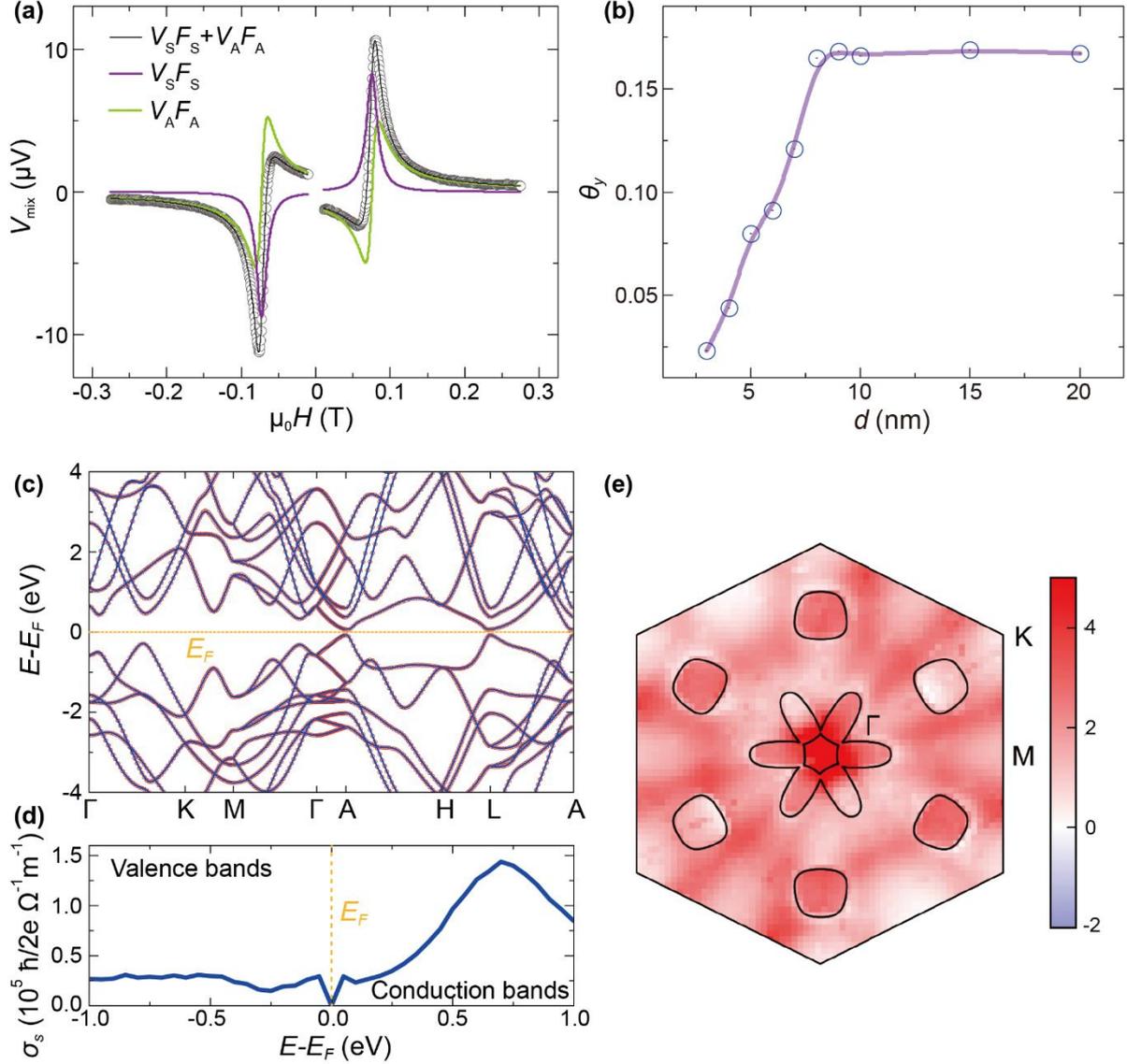

FIG. 1. Spin Hall conductivity of SnTe. (a) Spin-torque ferromagnetic resonance (ST-FMR) data from SnTe (8 nm)/Py at 7 GHz. (b) SnTe thickness ($d$) dependent spin Hall angle ($\theta_y$) from SnTe ($d$)/Py. (c) Comparison of SnTe band structures from *ab initio* calculation (red circles) with the band structure interpolated using maximally localized Wannier functions (blue lines). The yellow dashed horizontal line indicates the Fermi level. (d) Variation of spin Hall conductivity (SHC) in bulk SnTe as a function of the energy level. The yellow dashed vertical line marks the calculated Fermi level. Note that the SHC value at $E - E_F = 0$ point is excluded due to the band gap, which results in no charge contribution to transport. (e) $k$-resolved spin Berry curvature of SnTe in a slice of the Brillouin zone ($k_z = 0$) at the energy position of $E = E_F - 0.5$ eV. The color represents the magnitude of spin Berry curvature. Black lines are the intersections of the shifted Fermi surface with the slice of Brillouin zone.

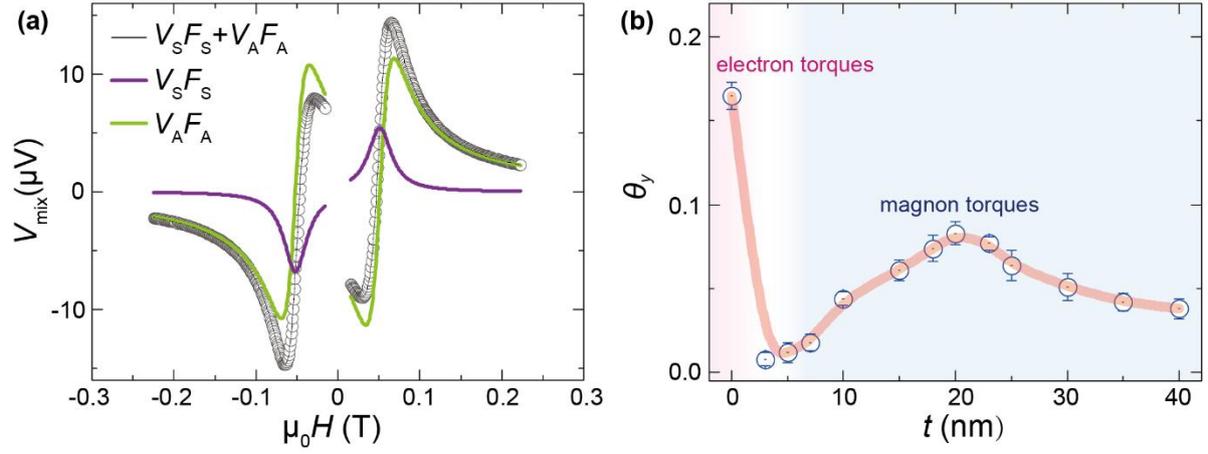

FIG. 2. Observation of magnon torques in SnTe/NiO/Py. (a) ST-FMR spectra from the SnTe (8 nm)/NiO (20 nm)/Py sample at 7 GHz. (b) NiO thickness ($t$) dependence of spin-orbit torque efficiency ($\theta_y$) in SnTe (8 nm)/NiO ($t$)/Py. Pink and blue shaded areas indicate electron- and magnon-mediated torques, respectively.

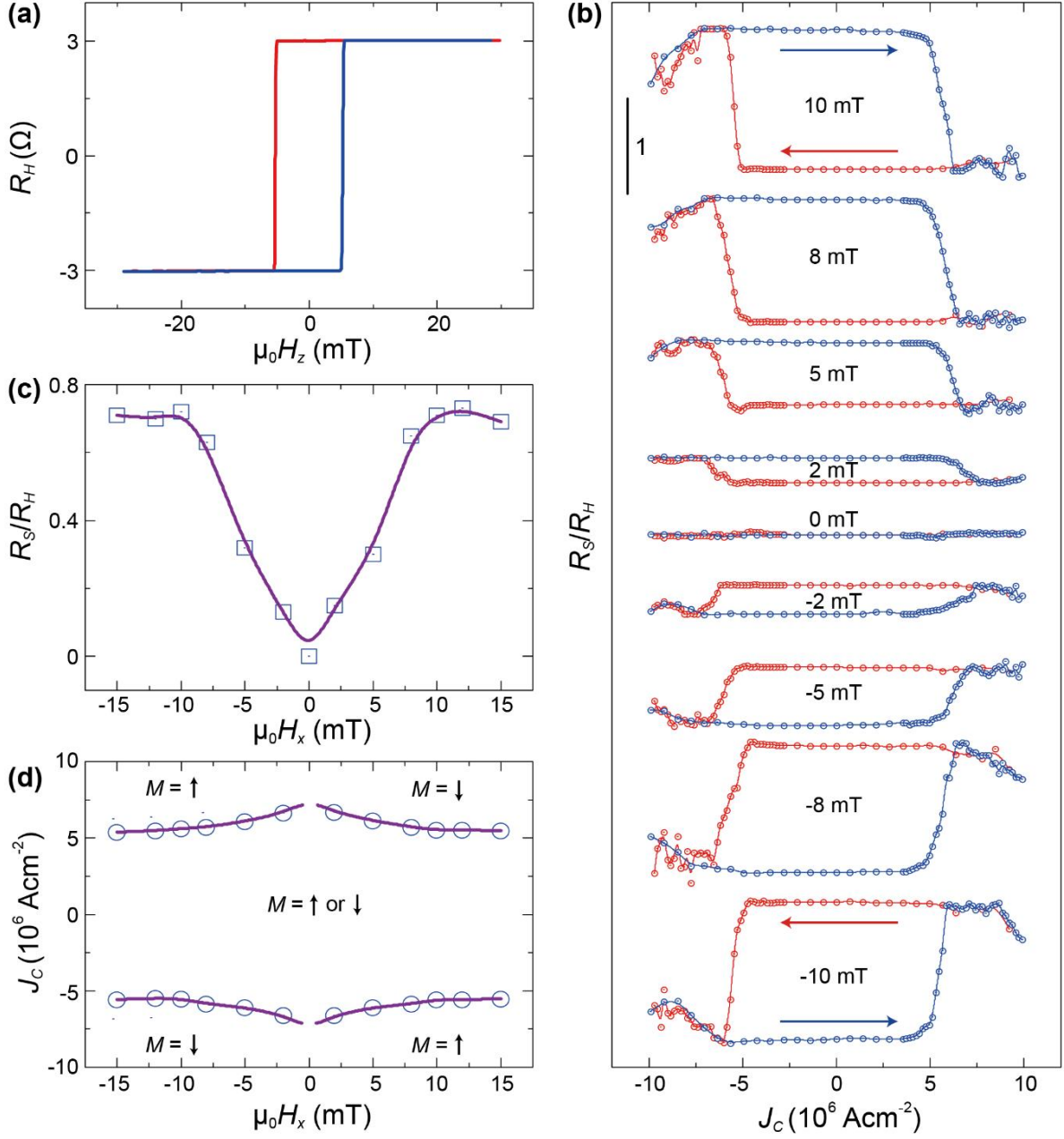

FIG. 3. Magnon torques-driven switching of perpendicular magnetization in SnTe (8 nm)/NiO (20 nm)/CoFeB. (a) Anomalous Hall loop measured by sweeping the out-of-plane magnetic field ($\mu_0H_z$). (b) Magnon-induced spin-torque switching under the in-plane magnetic field ($\mu_0H_x$) from −10 mT to 10 mT. (c) $\mu_0H_x$ dependence of the switching ratio ($R_S/R_H$). (d) $\mu_0H_x$ dependence of the critical switching current density ($J_C$).

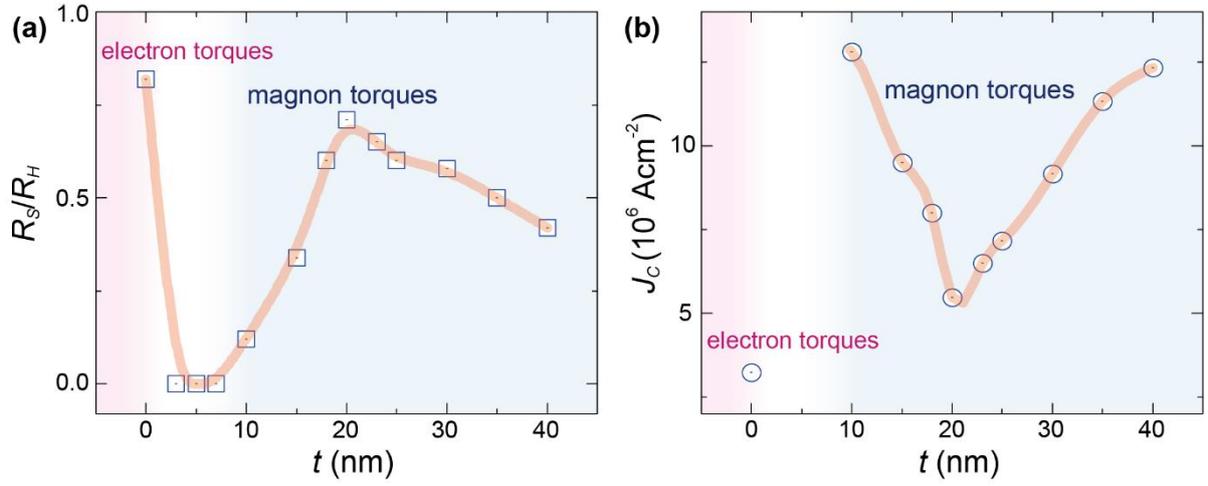

FIG. 4. NiO thickness dependence of spin-orbit torques in SnTe (8 nm)/NiO (*t*)/CoFeB. (a) *t* dependence of $R_S/R_H$. (b) *t* dependence of $J_C$. Pink and blue shaded areas indicate electron- and magnon-mediated torques, respectively. The discontinuity in (b) is due to no-switching.

Supplemental Material

# Magnon-mediated perpendicular magnetization switching by topological crystalline insulator SnTe with high spin Hall conductivity

Pengnan Zhao[1,#], Guoyi Shi[2,#], Wentian Lu[1,#], Lihuan Yang[1], Hui Ru Tan[3], Kaiwei Guo[1], Jia-Min Lai[1], Zhonghai Yu[1], Anjan Soumyanarayanan[3,4], Zhe Yuan[5,6], Fei Wang[1]*, Xiaohong Xu[1,†], and Hyunsoo Yang[2,‡]

[1]Key Laboratory of Magnetic Molecules and Magnetic Information Materials of Ministry of Education & School of Chemistry and Materials Science of Shanxi Normal University, Taiyuan 030006, China

[2]Department of Electrical and Computer Engineering, National University of Singapore, Singapore, 117576, Singapore

[3]Institute of Materials Research & Engineering, Agency for Science, Technology & Research (A*STAR), Singapore 138634, Singapore

[4]Department of Physics, National University of Singapore, Singapore 117551, Singapore

[5]Institute for Nanoelectronic Devices and Quantum Computing, Fudan University, Shanghai 200433, China

[6]Interdisciplinary Center for Theoretical Physics and Information Sciences, Fudan University, Shanghai 200433, China

*Contact author: feiwang.imr@gmail.com

†Contact author: xuxh@sxnu.edu.cn

‡Contact author: eleyang@nus.edu.sg

**Content**

**Section S1: Materials characterizations of SnTe/NiO films**

The crystalline quality of SnTe films is first characterized using *in-situ* reflection high-energy electron diffraction (RHEED) patterns. Figure S1 presents the RHEED patterns of an 8 nm thick SnTe film, which display long stripe shapes with discernible modulation, indicating a smooth surface morphology. The Raman spectrum of the 8 nm SnTe film is shown in Figure S2, revealing three dominant Raman-active peaks at 93, 121 and 139 cm$^{-1}$, which confirm the formation of SnTe[42]. Figure S3 displays the X-ray diffraction (XRD) image for the 8 nm SnTe film grown on $Al_2O_3$ (001) substrates, where the presence of the (222) reflection peak of SnTe confirms its epitaxial growth on the $Al_2O_3$ (001) substrates.

Figure S4 illustrates the transmission electron microscopy (TEM) results for a SnTe (6 nm)/NiO (20 nm)/Ti (2 nm)/$Co_{0.2}Fe_{0.6}B_{0.2}$ (0.9 nm)/MgO (2 nm)/Ta (1.5 nm) heterostructure grown on an $Al_2O_3$ (001) substrate. The TEM images show clear and well-defined interfaces (marked by a white dotted line), indicating no significant elemental intermixing. Moreover, the NiO layer exhibits polycrystalline morphological textures. The high-angle annular dark field (HAADF) scanning TEM image and corresponding energy dispersive X-ray (EDX) mappings are presented in Figure S4b, providing additional evidence of the high-quality interfaces in the heterostructure. We note the NiO layer shows polycrystalline morphological textures. The polycrystalline nature of NiO layer is also confirmed by the XRD analysis (Figure S5), displaying peaks at (111) and (220).

The Hall trace of an 8 nm SnTe film measured at 300 K is shown in Figure S6b. The positive slope of the linear Hall trace indicates that the SnTe film is dominated by hole-type carriers[26], and the chemical potential lies in valance bands. Figure S6c depicts the temperature ($T$) dependence of the longitudinal resistivity ($\rho_{xx}$) of the 8 nm SnTe film. $\rho_{xx}$ decreases with decreasing $T$, indicating its metallic behavior. Figure S6d summaries the dependence of $\rho_{xx}$ on the SnTe thickness ($d$) at 300 K. As the thickness increases, $\rho_{xx}$ decreases and eventually reaches a constant value.

We investigated the $T$ dependence of $\rho_{xx}$ in SnTe films with varying $d$, as shown in Figure S7. With the exception of 3 nm and 4 nm samples, all films demonstrate a metallic behavior, evidenced by a decrease in $\rho_{xx}$ with decreasing $T$. In topological (crystalline) insulators, the conductivity is predominantly governed by gapless topological surface states. As the film thickness decreases, the coupling between the upper and lower surface states leads to the opening of a gap. The band gap increases with decreasing the film thickness, leading to an increase in the film resistivity.

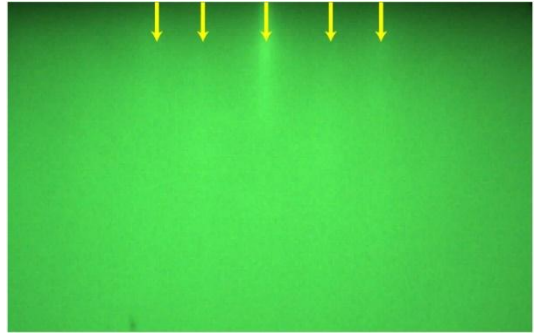

FIG. S1. Reflection high-energy electron diffraction (RHEED) patterns of an 8 nm SnTe film.

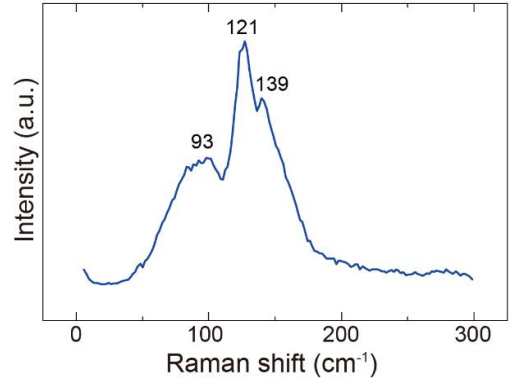

FIG. S2. Raman measurements of an 8 nm SnTe film.

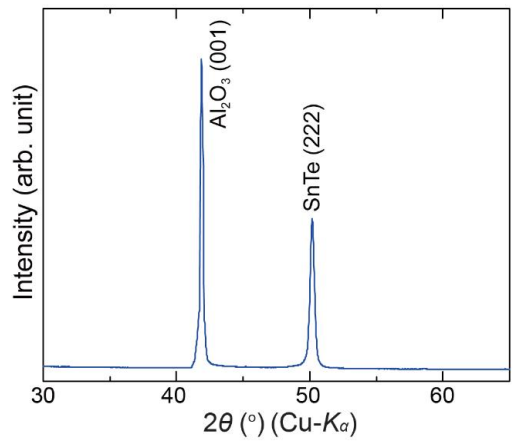

FIG. S3. X-ray diffraction (XRD) spectroscopy of an 8 nm SnTe film grown on a Al$_2$O$_3$ (001) substrate.

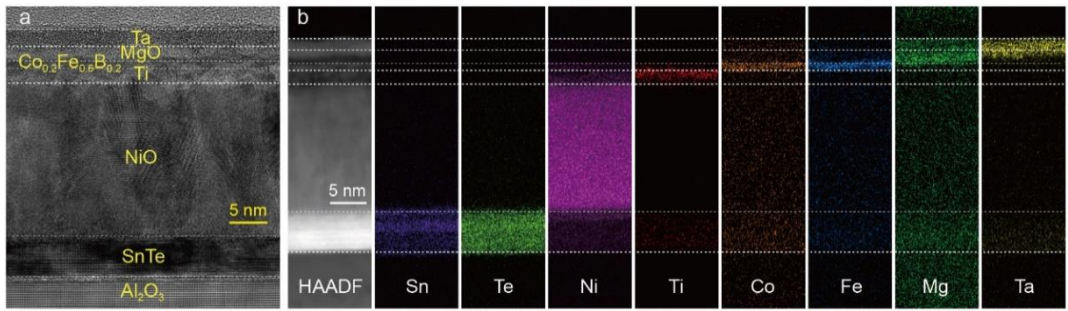

FIG. S4. Transmission electron microscopy (TEM) characterization of SnTe (6 nm)/NiO (20 nm)/Ti (2 nm)/Co$_{0.2}$Fe$_{0.6}$B$_{0.2}$ (0.9 nm)/MgO (2 nm)/Ta (1.5 nm). (a) Cross-sectional TEM result. (b) High-angle annular dark field (HAADF) scanning TEM image and corresponding energy dispersive X-ray (EDX) mappings of Sn, Te, Ni, Ti, Co, Fe, Mg, and Ta.

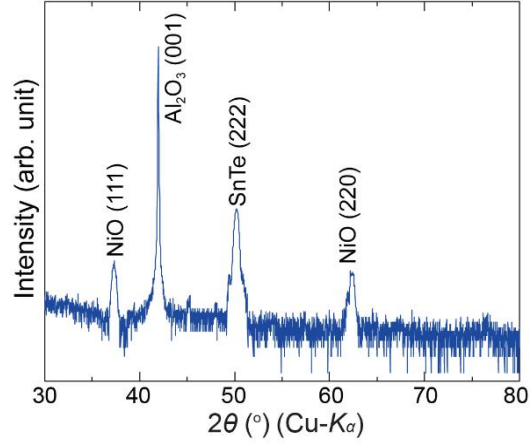

FIG. S5. XRD spectroscopy of an SnTe (8 nm)/NiO (50 nm) bilayer grown on a Al$_2$O$_3$ (001) substrate.

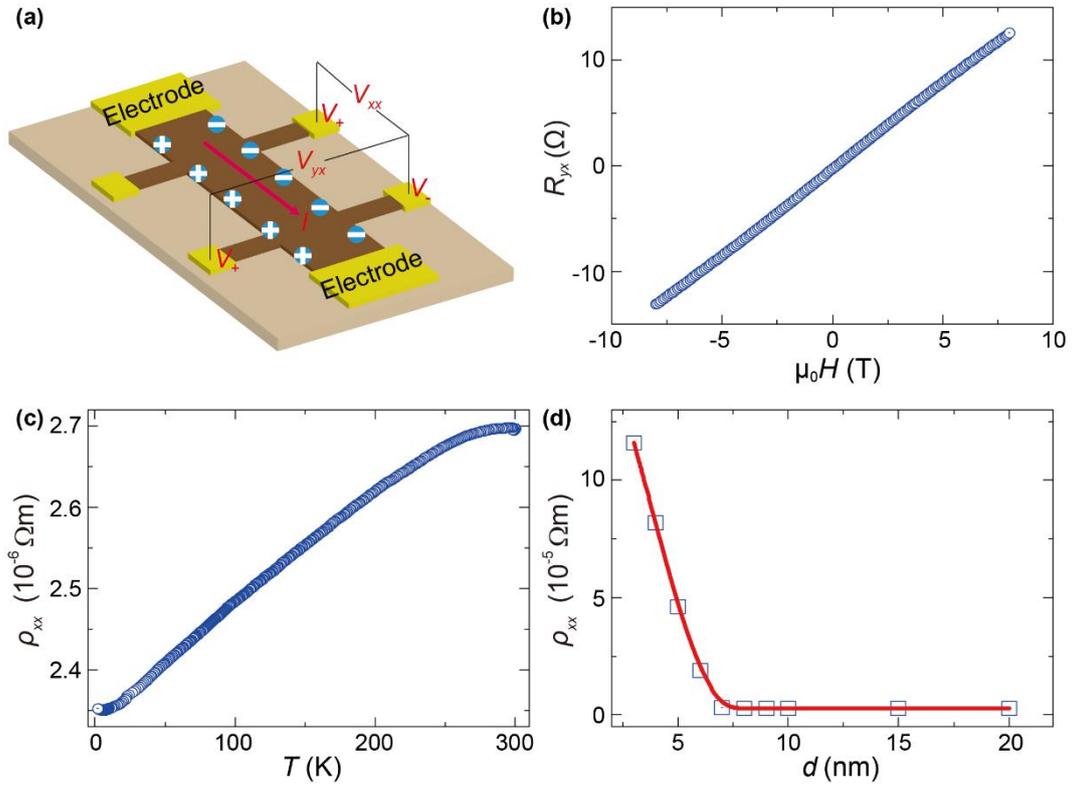

FIG. S6. Transport measurements of SnTe films. (a) Schematic illustration of the longitudinal (Hall) voltages $V_{xx}$ ($V_{yx}$) measurements in a Hall bar device. $I$ is the applied current. ⊕ and ⊖ are hole and electron carriers, respectively. (b) Hall trace of 8 nm SnTe film measured at 300 K. (c) Temperature ($T$) dependence of the longitudinal resistivity ($\rho_{xx}$) of an 8 nm SnTe film. (d) SnTe thickness ($d$) dependence of $\rho_{xx}$ at 300 K.

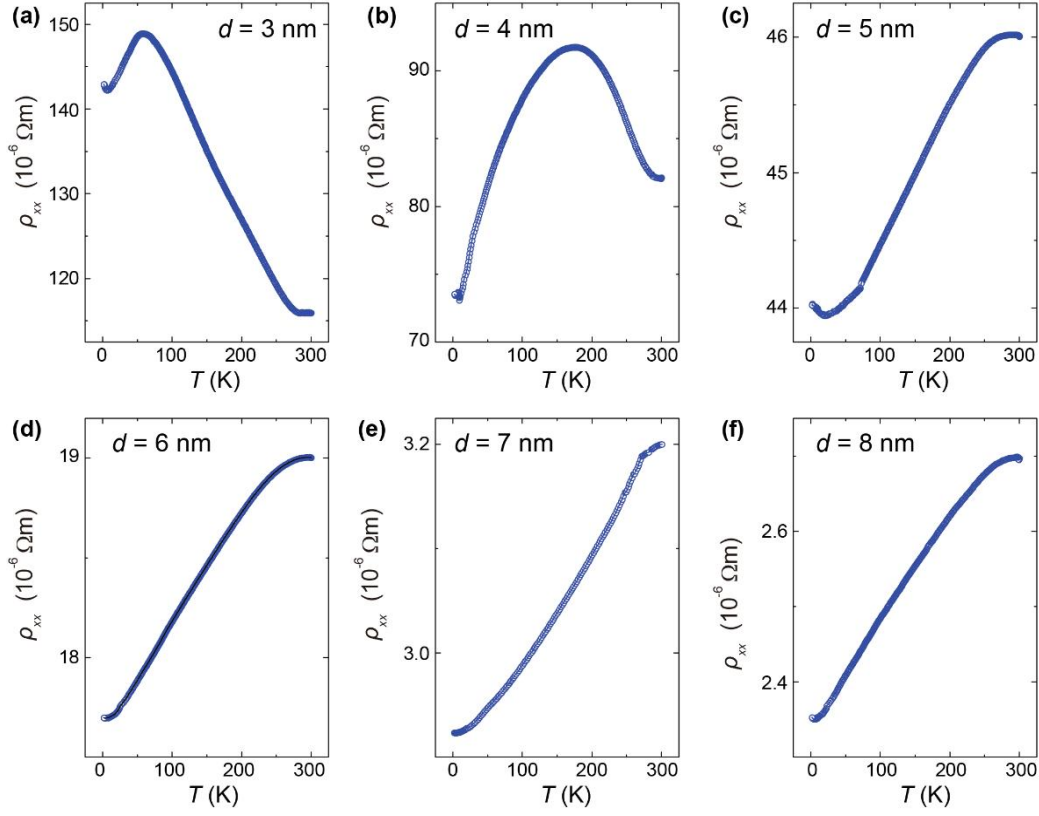

FIG. S7. $T$ dependence of $\rho_{xx}$ of different thicknesses of SnTe. (a) 3 nm. (b) 4 nm. (c) 5 nm. (d) 6 nm. (e) 7 nm. (f) 8 nm.

**Section S2: Spin-torque ferromagnetic resonance measurements of SnTe/NiO/Py**

The spin-torque ferromagnetic resonance (ST-FMR) linewidth of SnTe (8 nm)/Py (Figure S8a) is narrower than that of SnTe (8 nm)/NiO (20 nm)/Py (Figure 2a in the main text) at frequency $f = 7$ GHz. The linewidth of ST-FMR is directly related to the magnetic damping of Py. The exchange coupling between NiO and Py introduces additional damping in Py, resulting in an increased ST-FMR linewidth. As the thickness of the NiO layer increases, the exchange coupling is enhanced, leading to a broader linewidth[43].

Figure S9 shows the dependence of the spin Hall angle ($\theta_y$) and spin Hall conductivity ($\sigma_s$) on the thickness ($d$) of SnTe in SnTe ($d$)/Py. As the $d$ increases, $\theta_y$ increases and approaches saturation (Figure S9a). This behavior is described by the equation $\theta_y = \theta_\infty [1-\text{sech}(t/l_s)]$, where $\theta_\infty$ represents the spin efficiencies at infinite $d$, and $l_s$ is the spin diffusion length. Through fitting, the $l_s$ of SnTe is determined to be 4.1 nm. In the SnTe/Py heterostructures, the quality of the SnTe film impacts not only the spin conversion and transport properties but also ferromagnetic properties of the Py layer. These variations may lead to slight differences in the $\theta_y$.

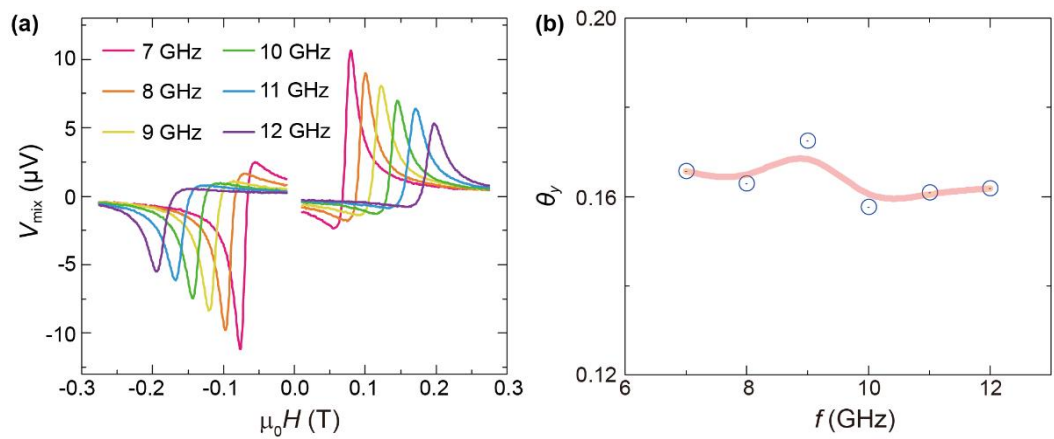

FIG. S8. Spin-torque ferromagnetic resonance (ST-FMR) results of a SnTe (8 nm)/Py sample. (a) ST-FMR spectra with frequencies ($f$) ranging from 7 to 12 GHz. (b) $f$ dependence of spin Hall angle ($\theta_y$).

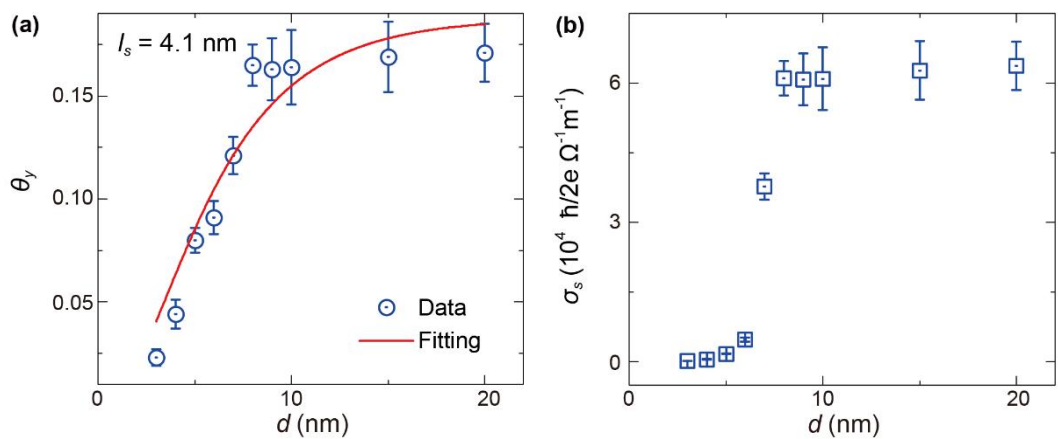

FIG. S9. Spin torque efficiencies and spin Hall conductivities of different SnTe thickness $d$. (a) Spin Hall angle $\theta_y$, the solid red line is fitting curve. (b) Spin Hall conductivity $\sigma_s$. The error bars reflect the standard deviation of 3 measurements.

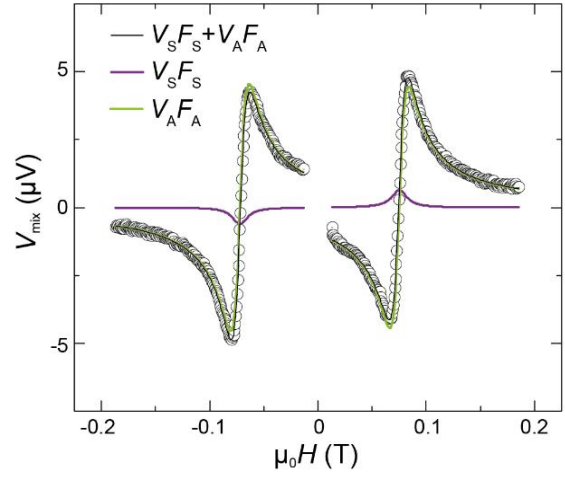

FIG. S10. ST-FMR signals of a SnTe (8 nm)/NiO (3 nm)/Py sample at 7 GHz.

## Section S3: Theoretical calculation results

**First-Principles Calculations:** *Ab initio* calculations are performed using the plane-wave ultrasoft pseudopotential method with a plane-wave basis set, as implemented in QUANTUM ESPRESSO[44-46]. We adopt an energy cutoff of 70 Ry and an electron density cutoff of 840 Ry for the plane-wave expansion, applying the generalized gradient approximation (GGA) for the exchange-correlation functional throughout the calculations[47-49]. The intrinsic spin Hall conductivity (SHC) is calculated using the maximally localized Wannier functions (MLWFs) method[50], implemented in Wannier90[31,51]. The intrinsic SHC ($\sigma_{\alpha\beta}^{\text{spin}\gamma}$), $\alpha, \beta, \gamma = x, y, z$, is computed using the Kubo-Greenwood formula[31,52].

$$\sigma_{\alpha\beta}^{\text{spin}\gamma}(\omega) = -\frac{e^2}{\hbar} \frac{1}{\Omega_c N_k} \sum_k \Omega_{\alpha\beta}^{\text{spin}\gamma}(\mathbf{k}) \tag{1}$$

where $\Omega_c$ is the cell volume, $N_k$ is the number of *k*-points used for sampling the Brillouin zone (BZ), $\Omega_{\alpha\beta}^{\text{spin}\gamma}(\mathbf{k})$ is the *k*-resolved Berry curvature term that sums the band-projected Berry curvature term $\Omega_{n,\alpha\beta}^{\text{spin}\gamma}(\mathbf{k})$ over occupied bands:

$$\Omega_{\alpha\beta}^{\text{spin}\gamma}(\mathbf{k}) = \sum_n f_{nk} \Omega_{n,\alpha\beta}^{\text{spin}\gamma}(\mathbf{k}) \tag{2}$$

where $f_{nk} = f(\varepsilon_{nk})$ is the Fermi-Dirac distribution function[31]. In the self-consistent field (SCF) calculation, we employ Monkhorst-Pack *k* meshes of 8×8×4. For the non-self-consistent field (NSCF) calculations used for constructing the MLWFs, we use *k* meshes of 12×12×6. The energy convergence criterion is set to $10^{-10}$ Ry. The unit cell contains three Sn atoms and Te atoms, and the atomic coordinators are fully relaxed with a force tolerance of $10^{-3}$ Ry/Bohr. For the construction of the MLWFs, we use an inner frozen window ranging from −0.5 to 12.5 eV and an outer disentanglement window from −0.5 to 17.5 eV, from which we extract 48 spinor Wannier functions. The spreads for both the disentanglement and Wannierization processes are converged to under $1\times10^{-10}$ Å$^2$. To compute $\sigma_{\alpha\beta}^{\text{spin}\gamma}$, we utilize a 200×200×100 *k*-point mesh, employing a 5×5×5 adaptive *k*-mesh refinement to accurately capture the rapid variation of spin Berry curvature in the BZ.

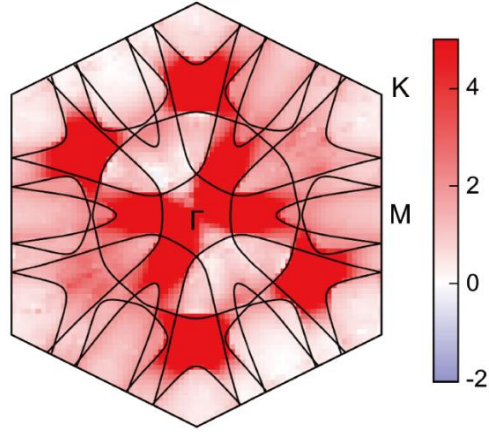

FIG. S11. *k*-resolved spin Berry curvature of SnTe in a slice of the Brillouin zone ($k_z = 0$) at the energy of $E = E_F + 0.7$ eV. The color represents the magnitude of spin Berry curvature. Black lines are the intersections of the shifted Fermi surface with the slice of Brillouin zone.

**Section S4: Hysteresis loop measurements of SnTe (8 nm)/NiO (t)/Py films**

We perform $T$-dependent hysteresis loop measurements on SnTe (8 nm)/NiO ($t$)/Py samples, varying the thickness ($t$) of the NiO layer. Figure S12a illustrates the hysteresis loop of the SnTe (8 nm)/NiO (3 nm)/Py film at 10 K. This is obtained after cooling from 380 to 10 K under a 3 T in-plane magnetic field. A noticeable shift of the hysteresis loop towards the negative magnetic field direction indicates the presence of an exchange bias effect. The exchange bias field ($\mu_0 H_{ex}$), calculated as the average of the negative and positive coercive fields, $\mu_0 H_{c1}$ and $\mu_0 H_{c2}$ respectively, is $-10.5$ mT at 10 K. This observation suggests the presence of antiferromagnetic ordering in the NiO layer. As the temperature increases, $\mu_0 H_{ex}$ diminishes and eventually vanishes at the blocking temperature ($T_b$), as shown in Figure S12b. Moreover, we examine hysteresis loops for different NiO thicknesses in the SnTe (8 nm)/NiO ($t$)/Py samples. It is observed that $T_b$ increases with thicker NiO layers, indicating a higher antiferromagnetic Néel temperature ($T_N$), summarized in Figure S12c. This trend is further supported by thickness-dependent coercive field ($\mu_0 H_c$) measurements conducted at room temperature, illustrated in Figure S12d. For the sample with $t = 3$ nm, $T_b$ is approximately 70 K, as depicted in Figures S12b and S12c. Notably, there is no significant increase in $\mu_0 H_c$ at room temperature (Figure S12d). This implies that the 3 nm NiO layer is not antiferromagnetic at room temperature and blocks electron-mediated spin currents. As the NiO thickness increases, the antiferromagnetic ordering gradually recovers, evidenced by the rise in both $T_b$ and $\mu_0 H_c$. However, as the spin transmission distance through NiO increases with its thickness, there is a concomitant increase in spin angular momentum loss. This balance between enhanced antiferromagnetic ordering and increased spin loss results in the optimal spin torque efficiency being achieved at an NiO thickness of $t = 20$ nm.

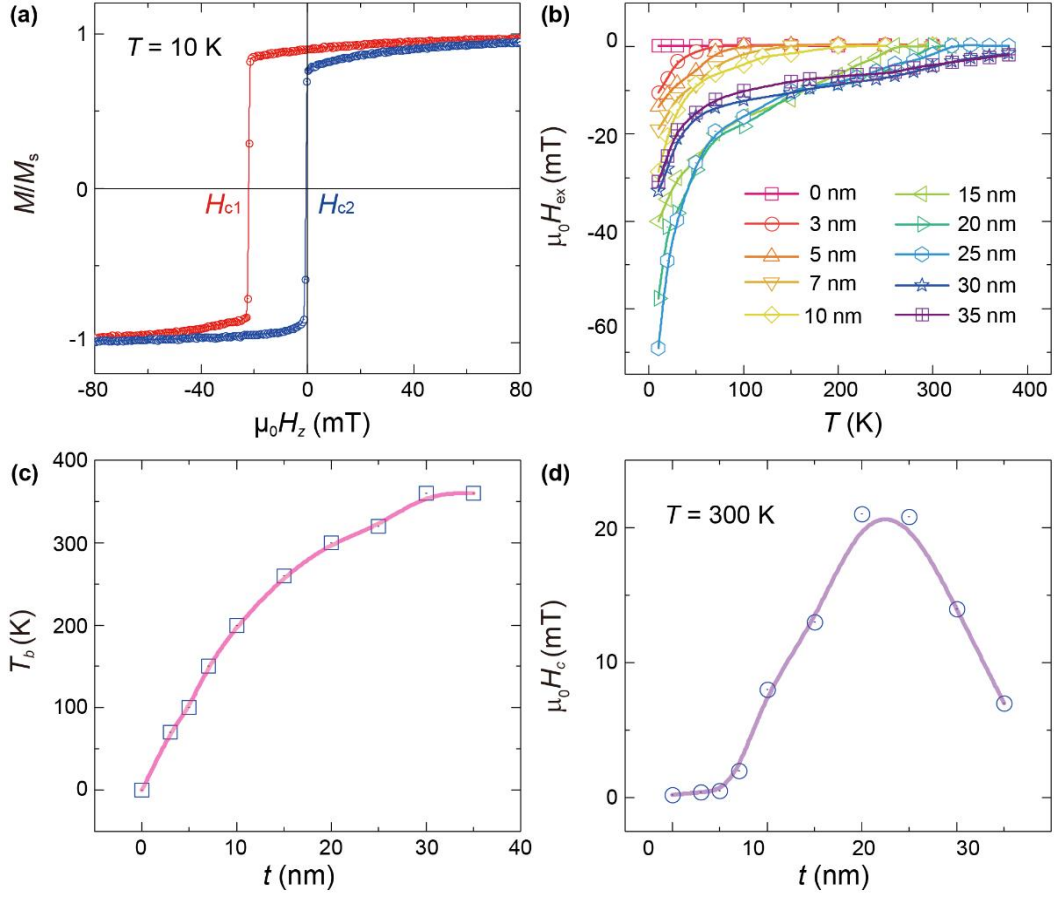

FIG. S12. Hysteresis loop measurements of SnTe (8 nm)/NiO ($t$)/Py. (a) Magnetic field ($\mu_0 H_z$) dependence of normalized magnetization at 10 K in the SnTe (8 nm)/NiO (3 nm)/Py samples. (b) Exchange bias field ($\mu_0 H_{ex}$) as a function of temperature ($T$) for various NiO thicknesses ($t$). (c) Blocking temperature ($T_b$) obtained from (b). (d) Coercive field ($\mu_0 H_c$) as a function of $t$ at room temperature.

# Section S5: Control samples of Bi$_2$Te$_3$/NiO/CoFeB

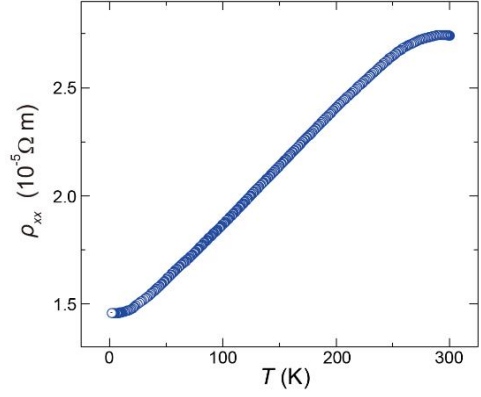

FIG. S13. $T$ dependent $\rho_{xx}$ of an 8 nm Bi$_2$Te$_3$ film.

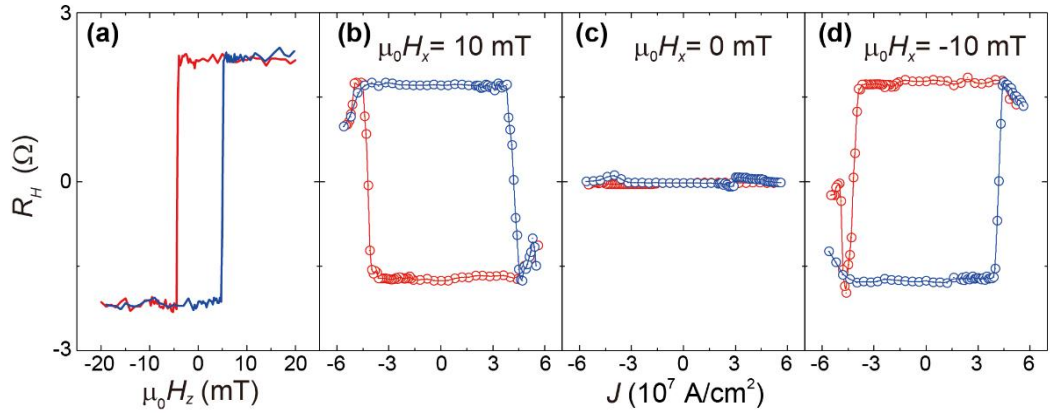

FIG. S14. Magnon-induced magnetization switching in the Bi$_2$Te$_3$ (8 nm)/NiO (25 nm)/CoFeB sample. (a) $\mu_0 H_z$ dependence of $R_H$. (b-d) Current-induced switching of the CoFeB layer under the in-plane magnetic field of $\mu_0 H_x$ = 10 mT (b), 0 mT (c), and -10 mT (d).

**Section S6: Calculation of power consumption of magnon devices**

We calculate the power consumption of magnon devices that consist of a spin source layer (SnTe and $Bi_2Te_3$) and a ferromagnetic layer (CoFeB), taking into account the current shunting into the CoFeB layer[36]. The Ti/CoFeB layer is 2.9 nm thick and consists of 2 nm Ti and 0.9 nm $Co_{20}Fe_{60}B_{20}$. The resistivity of the 2.9 nm Ti/CoFeB, 8 nm SnTe and 8 nm $Bi_2Te_3$ are 227.7, 270, and 2700 μΩcm, respectively. The power consumption is calculated by $P = I_{FM}^2 \rho_{FM} \frac{L}{WD_1} + I_S^2 \rho_S \frac{L}{WD_2}$, where $I_{FM}$ and $I_S$ are the respective currents flowing through the CoFeB and spin source layer. The calculated power consumption of the SnTe (8 nm)/NiO (20 nm)/CoFeB device is 22 times lower than that of the $Bi_2Te_3$ (8 nm)/NiO (25 nm)/CoFeB device.

We calculate the power consumption of the SnTe(8 nm)/CoFeB sample, without the NiO layer. The results indicate that the power consumption for electron-mediated torques in SnTe/CoFeB is 2.8 times lower than that for magnon torques in SnTe/NiO/CoFeB (Table I).

TABLE I. Device parameters of SnTe/CoFeB, SnTe/NiO/CoFeB and $Bi_2Te_3$/NiO/CoFeB. All switching measurements are conducted with an external magnetic field of 10 mT.

| Materials | $I_S$ (mA) | $I_{FM}$ (mA) | $\rho_S$ (μΩcm) | $P$ (mW) |
|---|---|---|---|---|
| SnTe (8 nm)/NiO (20 nm)/CoFeB | 8.8 | 3.783 | 270 | 74.74 |
| $Bi_2Te_3$ (8 nm)/NiO (25 nm)/CoFeB | 6.8 | 29.24 | 2700 | 1654.4 |
| SnTe (8 nm)/CoFeB | 5.2 | 2.24 | 270 | 26.13 |

**Section S7: Evaluation of temperature rise during magnetization switching**

To assess the heating effect during magnon torque switching measurements, we measured the *T*-dependent resistance change, following the approach used in previous studies[19,53]. As shown in Figure S15, at the current density applied during the switching measurements, the temperature increase in the SnTe (8 nm)/NiO (20 nm)/CoFeB sample is measured to be 11.3 K.

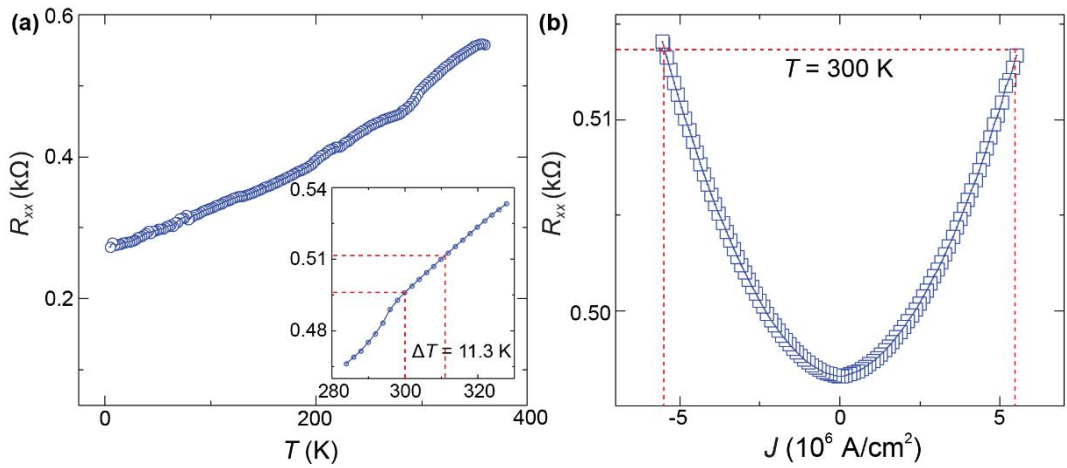

FIG. S15. Evaluation of temperature rise during SOT switching measurements of the SnTe (8 nm)/NiO (20 nm)/CoFeB sample. (a) *T* dependence of longitudinal resistance ($R_{xx}$). The inset shows enlarged version of the $R_{xx}$-*T* curve from 280 to 330 K. (b) Current density (*J*) dependence of $R_{xx}$ at 300 K.

**Section S8: Magnetization switching of SnTe/NiO/CoFeB**

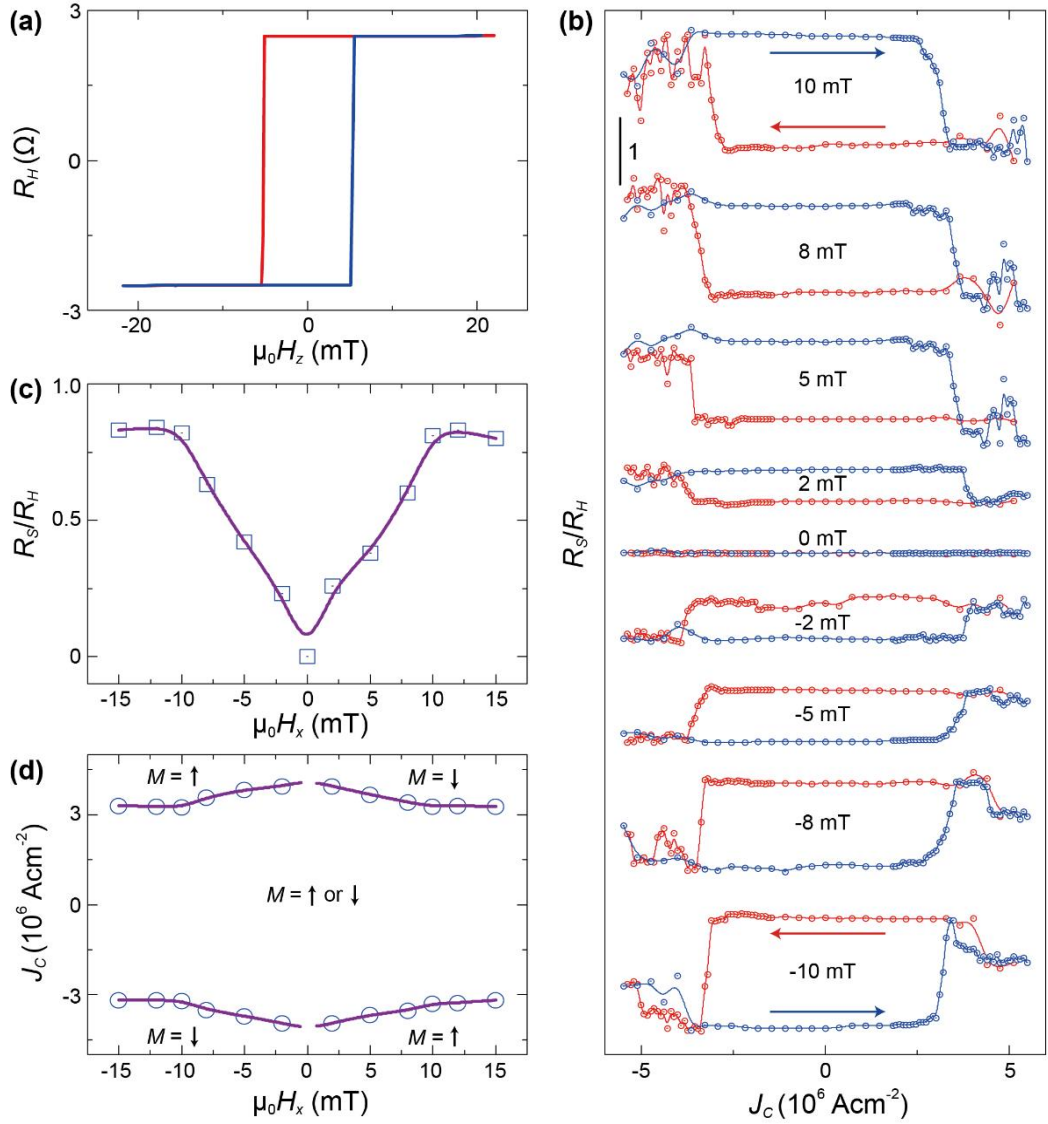

FIG. S16. Electron torques-driven switching of perpendicular magnetization in SnTe (8 nm)/CoFeB. (a) Anomalous Hall curve measured by sweeping $\mu_0H_z$. (b) Magnon-induced spin-torque switching under $\mu_0H_x$ from −10 mT to 10 mT. (c) $\mu_0H_x$ dependence of switching ratio ($R_S/R_H$). (d) $\mu_0H_x$ dependence of critical switching current density ($J_C$).

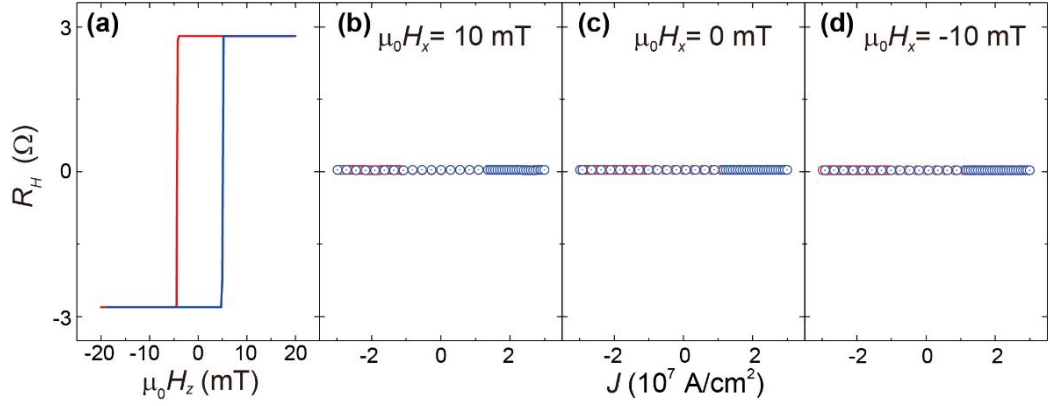

FIG. S17. Current-driven switching of perpendicular magnetization in SnTe (8 nm)/NiO (3 nm)/CoFeB. (a) $\mu_0 H_z$ dependence of $R_H$. (b-d) Current-induced switching of the CoFeB layer under $\mu_0 H_x$ = 10 mT (b), 0 mT (c), and -10 mT (d).

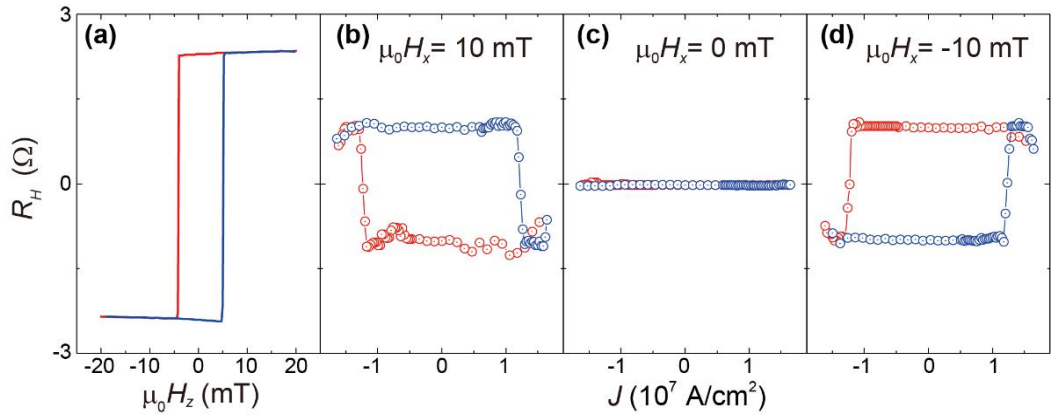

FIG. S18. Magnon-induced magnetization switching in the SnTe (8 nm)/NiO (40 nm)/CoFeB sample. (a) $\mu_0 H_z$ dependence of $R_H$. (b-d) Current-induced switching of the CoFeB layer under $\mu_0 H_x$ = 10 mT (b), 0 mT (c), and -10 mT (d).

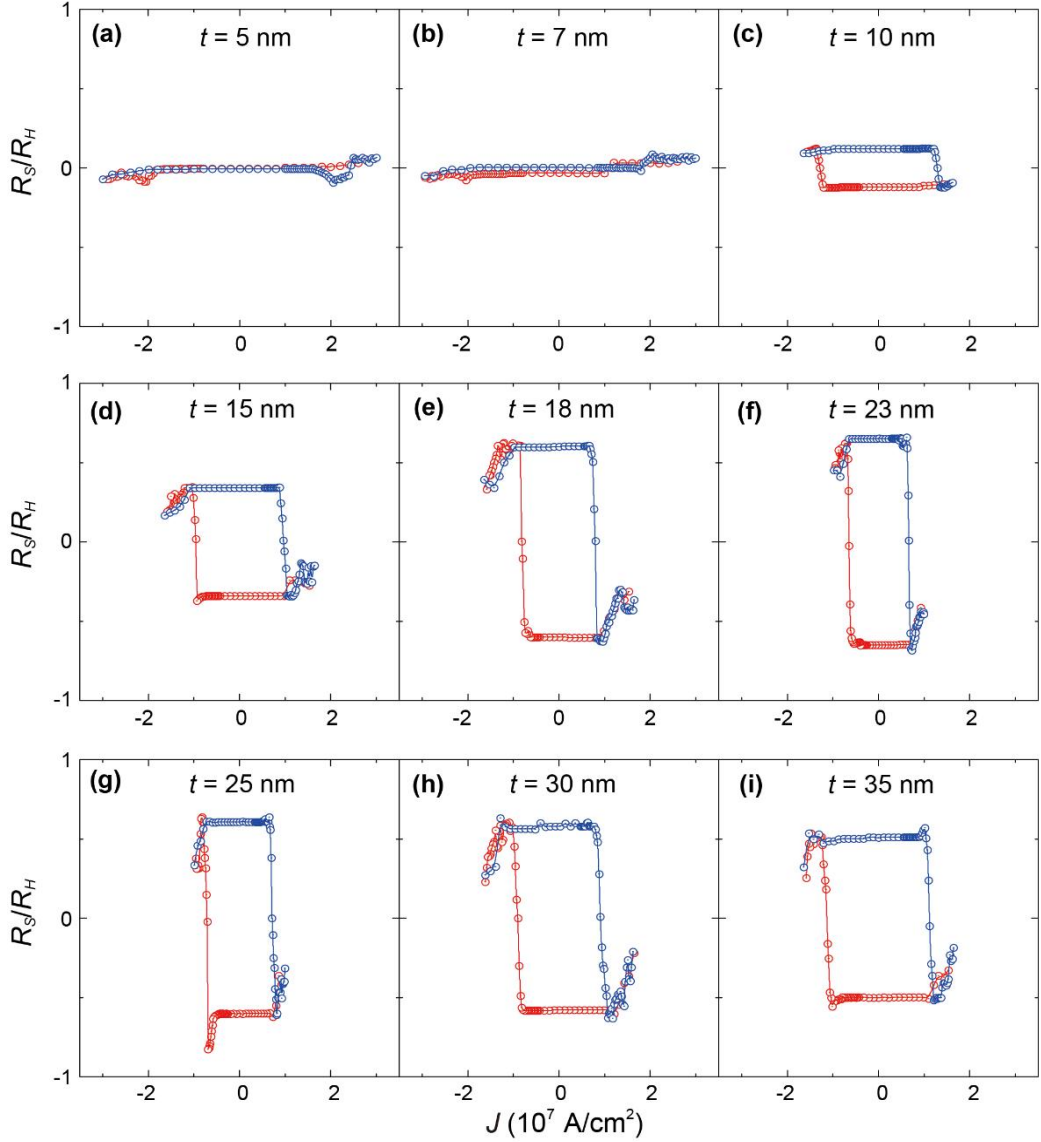

FIG. S19. Current-induced magnetization switching in SnTe (8 nm)/NiO ($t$)/CoFeB samples (a) $t$ = 5 nm. (b) $t$ = 7 nm. (c) $t$ = 10 nm. (d) $t$ = 15 nm. (e) $t$ = 18 nm. (f) $t$ =23 nm. (g) $t$ = 25 nm. (h) $t$ = 30 nm. (i) $t$ = 35 nm.

## Section S9: Magnetization switching of SnTe/Cu/CoFeB and SnTe/MgO/CoFeB

We conducted magnetization switching experiments on the control samples SnTe (8 nm)/Cu (3 nm)/CoFeB, with a light metal Cu insertion, and SnTe (8 nm)/MgO (3 nm)/CoFeB, with an insulator MgO insertion. Clear magnetization switching is observed in the SnTe (8 nm)/Cu (3 nm)/CoFeB sample (Figure S20); in contrast, switching is absent in the SnTe (8 nm)/MgO (3 nm)/CoFeB sample (Figure S21). These observations indicate that a few nanometers of NiO are similar to the nonmagnetic insulator MgO, which effectively suppress magnon transports.

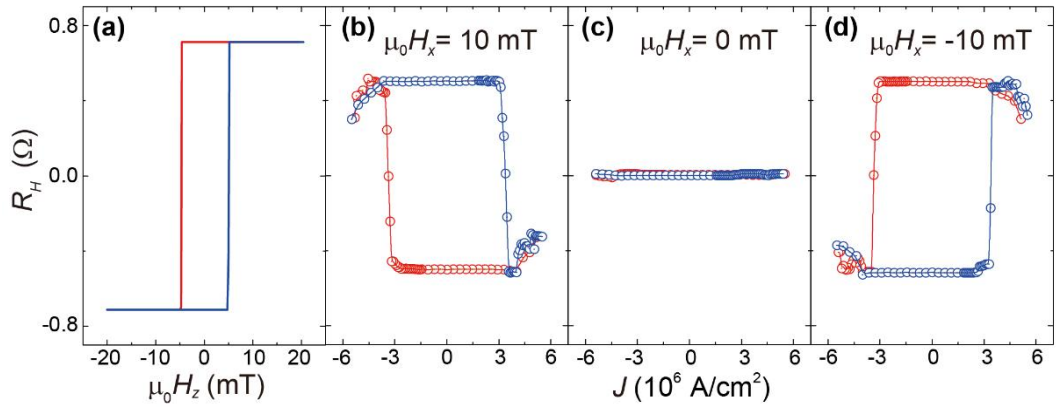

FIG. S20. Current-driven switching of perpendicular magnetization in SnTe (8 nm)/Cu (3 nm)/CoFeB. (a) $\mu_0 H_z$ dependence of $R_H$. (b-d) Current-induced switching of the CoFeB layer under $\mu_0 H_x$ = 10 mT (b), 0 mT (c), and -10 mT (d).

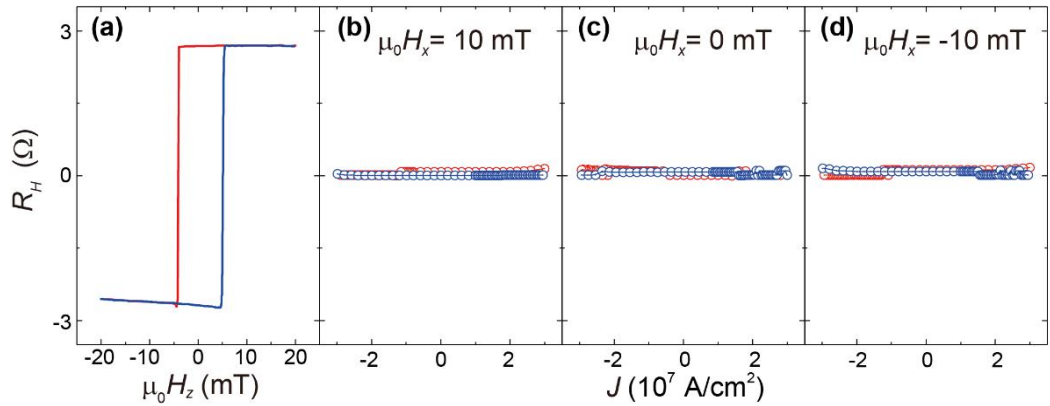

FIG. S21. Current-driven switching of perpendicular magnetization in SnTe (8 nm)/MgO (3 nm)/CoFeB. (a) $\mu_0 H_z$ dependence of $R_H$. (b-d) Current-induced switching of the CoFeB layer under $\mu_0 H_x$ = 10 mT (b), 0 mT (c), and -10 mT (d).

**Section S10: Comparison of topological material-based and Pt-based magnon devices**

We note that for Pt-based samples, there is a clear increase in the magnon efficiency when the NiO thickness is 0.5-2 nm, followed by a rapid decay with increasing the NiO thickness[11,18,37,38]. In contrast, for topological materials such as $Bi_2Se_3$, $Bi_2Te_3$ and SnTe, the spin-orbit efficiency is essentially zero when the NiO thickness is 1-3 nm, with maximum magnon efficiency appearing when the NiO thickness reaches around 20-25 nm[11,17]. This discrepancy can be attributed to the impact of the underlying spin source layer materials on the anisotropy and exchange interaction in NiO, leading to modifications in spin transport properties.

We systematically investigated Pt (6 nm)/NiO (*d*)/Co (3 nm) and $Bi_2Te_3$ (6 nm)/NiO (*d*)/Co (3 nm) samples[11]. We observed a slight increase in the spin-to-charge conversion efficiency for Pt (6 nm)/NiO (*d*)/Co (3 nm) samples at *d* ~ 1 nm, compared to that at *d* = 0 nm, consistent with the previous findings[37,38]. However, this increase at *d* ~ 1 nm was not observed in $Bi_2Te_3$ (6 nm)/NiO (*d*)/Co (3 nm) samples. To investigate further, we conducted XRD and magnetic property measurements for both sample types, as shown in Extended Data Figure 8 of Reference 19. Our analysis revealed that the slight increase observed in Pt-based samples is primarily due to improved NiO crystalline texturing, resulting from better lattice matching between NiO and Pt compared to that between NiO and $Bi_2Te_3$. The lower crystalline texturing of NiO in $Bi_2Te_3$-based samples suppresses antiferromagnetic magnons and spin fluctuation when the thickness of NiO is thin (e.g. 1 nm).

We now explain the formation of a peak in magnon efficiency at a particular NiO thickness of 20-25 nm in topological material-based samples. In the three-layer structure of spin source layer/NiO/ferromagnet, the magnon current is initiated by the charge current in the spin source layer. Insufficient charge-spin conversion efficiency in the spin source layer limits magnons excitation, resulting in a weak magnon current that may not be enough to transmit the thick NiO layer for magnetization switching. This limitation becomes evident in the significant drop in the magnon efficiency as the NiO thickness increases, particularly when Pt serves as the spin source layer. Notably, using topological materials as spin source layers can mitigate this decay due to their strong charge-spin conversion capabilities. When topological materials are employed, a robust spin accumulation at the topological material/NiO interface stimulates a considerable magnon current in NiO. This excitation and transport are intricately linked to the antiferromagnetic order of NiO, gradually increasing with the formation of NiO antiferromagnetic order. Nonetheless, an excessively thick NiO layer promotes the spin angular momentum loss during magnon transmission, resulting in a reduced magnon efficiency. Therefore, the NiO thickness for the maximum magnon efficiency, e.g., $Bi_2Se_3$/NiO/Py at 25 nm,[17] $Bi_2Te_3$/NiO/Co at 15-20 nm,[11] and our SnTe/NiO/Py at 20 nm, is predominantly determined by the underlying spin source layer material.

**Section S11: Summary of magnetization switching by magnon torques**

Table II summaries magnetization switching by magnon torques. Only a limited number of systems have demonstrated the ability to achieve magnetization switching via magnon torques, and most of these systems are constrained to low-temperature observations, with few exceptions for topological insulator-based samples. The primary limitation arises from the fact that, in the three-layer structure of spin source layer/antiferromagnetic insulator/ferromagnet, the magnon current is excited through spin accumulation at the spin source layer/antiferromagnet interface. Effective magnon excitation requires the spin source material to possess a sufficiently large spin torque efficiency to produce adequate spin accumulation, particularly at room temperature. Consequently, the selection of suitable spin source materials becomes significantly constrained.

Moreover, we grow (001)-oriented $Bi_{0.9}Sb_{0.1}$ on $Al_2O_3$ (001) substrates using molecular beam epitaxy, and RHEED, XRD, and atomic force microscopy (AFM) confirmed its high crystalline quality and smooth surface (Figure S22). After further deposition of 20 nm NiO and CoFeB, we do not observe perpendicular magnetization switching (Figure S23), likely due to the lower spin-orbit torque efficiency of the (001)-oriented $Bi_{0.9}Sb_{0.1}$ compared to (012)-oriented samples[54]. Note that the growth of (012)-oriented $Bi_{0.9}Sb_{0.1}$ requires a ferromagnetic layer of $Mn_{0.6}Ga_{0.4}$ as a buffer layer, which hinders our experimental observations of CoFeB layer switching. Moreover, no magnon torque-driven switching is detected in the Pt (8 nm)/NiO (20 nm)/CoFeB structure at room temperature, as shown in Figure S24.

TABLE II. Summary of magnetization switching by magnon torques.

| Magnon torque system | Highest observed temperature | References |
|---|---|---|
| $Bi_2Se_3$/NiO/NiFe | **300 K** | [17] |
| YIG/NiO/Pt | 150K | [18] |
| $SrRuO_3$/NiO/$SrIrO_3$ | 60K | [19] |
| $Bi_2Te_3$/NiO/CoFeB | **300K** | [21] |
| $SrRuO_3$/$BiFeO_3$/$SrIrO_3$ | 70K | [22] |
| SrRuO$_3$/LaMnO$_3$/SrIrO$_3$ | 70K | [55] |
| SnTe/NiO/CoFeB | **300K** | **Our work** |

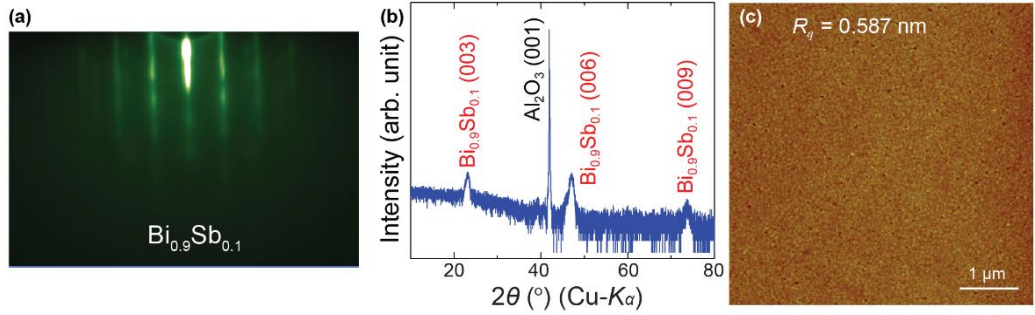

FIG. S22. Structural characterizations of an 8 nm Bi$_{0.9}$Sb$_{0.1}$ film grown on Al$_2$O$_3$ (001) substrate. (a) RHEED patterns. (b) XRD spectroscopy. (c) Atomic force microscopy (AFM) image with root mean-squared roughness ($R_q$) of 0.587 nm.

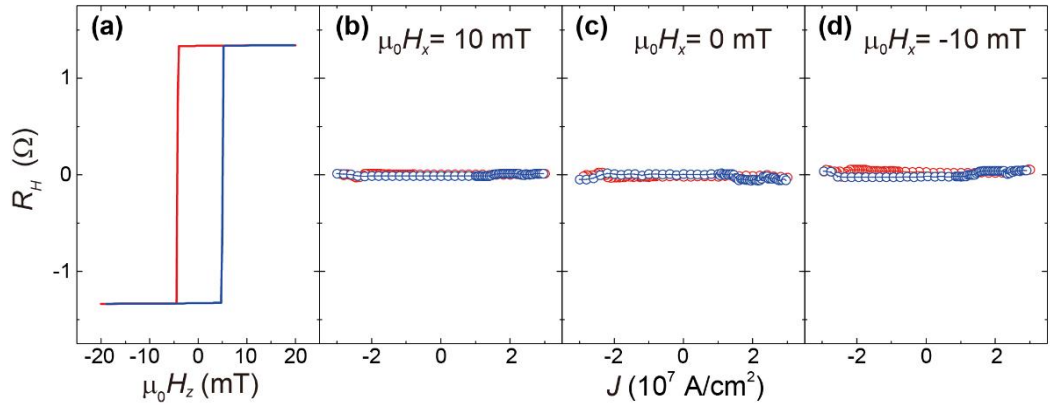

FIG. S23. Current-driven switching of perpendicular magnetization in Bi$_{0.9}$Sb$_{0.1}$ (8 nm)/NiO (20 nm)/CoFeB. (a) $\mu_0 H_z$ dependence of $R_H$. (b-d) Current-induced switching of the CoFeB layer under $\mu_0 H_x$ = 10 mT (b), 0 mT (c), and -10 mT (d).

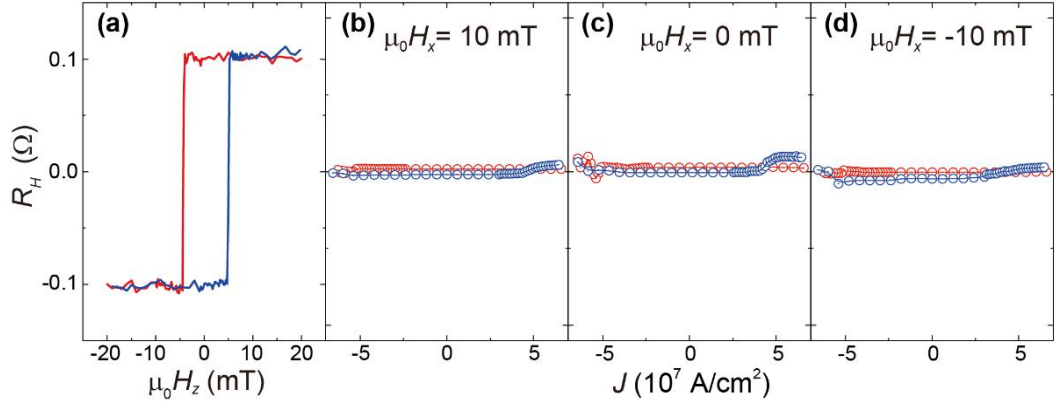

FIG. S24. Current-driven switching of perpendicular magnetization in Pt (8 nm)/NiO (20 nm)/CoFeB. (a) $\mu_0H_z$ dependence of $R_H$. (b-d) Current-induced switching of the CoFeB layer under $\mu_0H_x$ = 10 mT (b), 0 mT (c), and -10 mT (d).